\documentclass[prl,reprint]{revtex4-1} 
 
\usepackage{graphicx} 
\usepackage{bm}% bold math 
\usepackage{hyperref}% add hypertext capabilities 
\usepackage{color} 
\usepackage{cases}
\begin{document} 
 
\title{Dynamic Response of Wigner Crystals} 
 
\author{Lili Zhao} 
\affiliation{International Center for Quantum Materials, 
Peking University, Haidian, Beijing 100871, China} 
\author{Wenlu Lin} 
\affiliation{International Center for Quantum Materials, 
Peking University, Haidian, Beijing 100871, China} 
\author{Yoon Jang Chung} 
\affiliation{Department of Electrical Engineering, 
Princeton University, Princeton, New Jersey 08544, USA} 
\author{Adbhut Gupta} 
\affiliation{Department of Electrical Engineering, 
Princeton University, Princeton, New Jersey 08544, USA}
\author{Kirk W. Baldwin} 
\affiliation{Department of Electrical Engineering, 
Princeton University, Princeton, New Jersey 08544, USA}
\author{Loren N. Pfeiffer} 
\affiliation{Department of Electrical Engineering, 
Princeton University, Princeton, New Jersey 08544, USA} 
\author{Yang Liu} 
\email{liuyang02@pku.edu.cn} 
\affiliation{International Center for Quantum Materials, 
  Peking University, Haidian, Beijing 100871, China} 
 
\begin{abstract} 
 
  The Wigner crystal, an ordered array of electrons, is one of the
  very first proposed many-body phases stabilized by the
  electron-electron interaction. This electron
  solid phase has been reported in ultra-clean two-dimensional
  electron systems at extremely low temperatures, where the Coulomb
  interaction dominants over the kinetic energy, disorder potential
  and thermal fluctuation. We closely examine this quantum phase with capacitance measurements where the device length-scale is comparable with the crystal's correlation length. The extraordinarily high performance of our technique makes it possible to quantitatively study the dynamic response of the Wigner crystal within the single crystal regime. Our result will greatly boost the study of this inscrutable electron solid.
         
\end{abstract} 
 
\pacs{} 
 
\maketitle 
 
Interacting two-dimensional electron system (2DES) subjected to high
perpendicular magnetic fields ($B$) and cooled to low temperatures
exhibits a plethora of exotic states \cite{Jain.CF.2007}. The Wigner
crystal (WC) \cite{Wigner.PR.1934} terminates the sequence of fractional
quantum Hall states at very small landau level filling factor  \cite{Jiang.PRL.1990, Goldman.PRL.1990, Li.PRL.1991, Santos.PRL.1992, Sajoto.PRL.1993, Pan.PRL.2002, Maryenko.NC.2018, Hossain.Pnas.2020, Chung.PRL.2022, Lozovik.JETP.1975, Lam.PRB.1984, Levesque.PRB.1984, Andrei.PRL.1988, Williams.PRL.1991, Li.PRL.1997, Ye.PRL.2002, Chen.PRL.2004, Li.SSC.1995, Deng.prl.2019, Chen.np.2006, Drichko.PRB.2016, Tiemann.NP.2014}. This electron solid is pinned by the ubiquitous
residual disorder, manifests as an insulating phase in DC
transport \cite{Jiang.PRL.1990, Goldman.PRL.1990, Li.PRL.1991, Santos.PRL.1992,
  Sajoto.PRL.1993, Pan.PRL.2002, Maryenko.NC.2018, Hossain.Pnas.2020, Chung.PRL.2022}, and the electrons'
collective motion is evidenced by a resonance in AC
transport \cite{Lozovik.JETP.1975, Lam.PRB.1984, Levesque.PRB.1984,
  Andrei.PRL.1988, Williams.PRL.1991, Li.PRL.1997, Ye.PRL.2002, Chen.PRL.2004}. A
series of experiments have been applied to investigate
this correlated solid, such as the nonlinear $I-V$ response
\cite{Goldman.PRL.1990, Williams.PRL.1991}, the noise spectrum
\cite{Li.PRL.1991}, the huge dielectric constant \cite{Li.SSC.1995},
the weak screening efficiency \cite{Deng.prl.2019}, the melting
process \cite{Chen.np.2006, Drichko.PRB.2016, Deng.prl.2019}, the nuclear magnetic resonance \cite{Tiemann.NP.2014} and the optics \cite{Zhou.Nature.2021, Smolenski.Nature.2021}. 
 
Capacitance measurements have revealed a series of quantum phenomena
\cite{Mosser.SSC.1986, Ashoori.PRL.1992, Smith.PRB.1986,
  Yang.PRL.1997, Eisenstein.PRB.1994, Zibrov.nature.2017,
  Irie.APE.2019, Eisenstein.PRL.1992, Jo.PRB.1993, Li.Science.2011,
  Zibrov.PRL.2018, Tomarken.PRL.2019, Deng.prl.2019}. In this work, we examine the
WC formed in an ultra-high mobility 2DES at
$\nu \lesssim$ 1/5 using high-precision capacitance measurement
\cite{Zhao.RSI.2022, Zhao.CPL.2022}. We find an exceedingly large
capacitance at low measurement frequency $f$ while the conductance is almost
zero. This phenomenon is inconsistent with transporting electrons, but rather an
evidence that the synchronous vibration of electrons induces a
polarization current. When we increase $f$, our high-precision measurement captures the fine structure of the
resonance response with a puzzling "half-dome" structure. Our systematic, quantitative
results provide an in-depth insight of this murky quantum phase.
 
\begin{figure*}[!htbp] 
\includegraphics[width=0.9\textwidth]{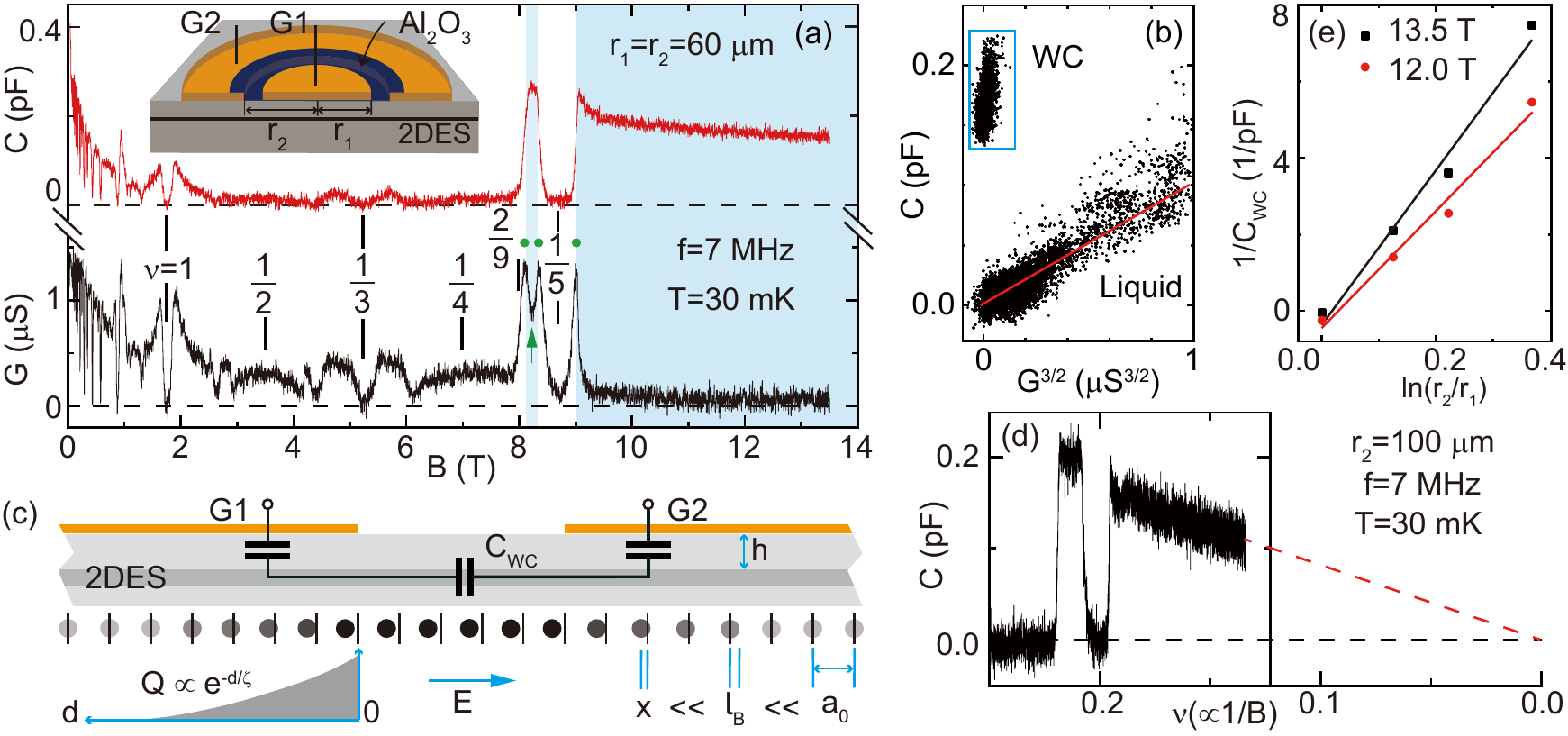} 
\caption{(color online) (a) $C$ and $G$ measured from the $r_1=r_2=$
	60 $\mu$m sample with 7 MHz excitation at 30 mK. The horizontal dashed
  lines represent the zeros of $C$ or $G$. The blue shaded regions
  mark the presence of WC. Inset is the cartoon of
  our device. (b) The correlation between $C$ and $G$ in panel (a)
  data. Transporting current dominates at $B<$
  8 T where $C\propto G^{3/2}$, indicated the red solid line. When the
  WC polarization current dominates, $C\simeq 0.2$ pF and $G$ is about
  zero (the blue box).
  (c) The schematic model describing the
  collective motion of electrons in the pinned WC. $h$ is the depth of 2DES. The equally spaced (by the lattice constant $a_0$)
  vertical bars represent the equilibrium position of
  electrons. The gray-scaled solid circles represent the
  electron position at finite external electric field
  $\mathbf{E}$. The darker gray corresponds to larger electron
  displacement $\mathbf{x}$. The radius of individual electron
  is about the magnetic length $l_{\text{B}}$. The
  accumulated charge $Q$ is proportional to
  $\nabla\cdot\mathbf{x}$, and decays exponentially as a function
  of the distance $d$ from the gate boundary. $\zeta$ is the
  decay length. $C_{\text{WC}}$ is the effective capacitance
  of WC in the un-gated region between the two
  gates.  (d) $C$ v.s. $\nu$ of the $r_2$=100 $\mu$m
  sample. The black dashed line is the zero of $C$. The red
  dashed line is the linear extension of data, showing that $C=0$ at the extreme quantum limit $\nu=0$. (e)
  $1/C_{\text{WC}}$ v.s. ln($r_2/r_1$) at two different
  magnetic field.
  }
\end{figure*} 

Our sample consists an ultra-clean low-density 2DES confined in a
70-nm-wide GaAs quantum well with electron density
$n\simeq4.4\times10^{10}$ cm$^{-2}$ and mobility $\mu\simeq $ 17
$\times10^6$ cm$^2$/(V$\cdot$s). Each device has a pair of front
concentric gates G1 and G2, whose outer and inner radius are $r_1$ and
$r_2$, respectively; see the inset of Fig. 1(a) \footnote{See Supplemental Material for detailed description of our sample information and measurement techniques.}. We study four devices
with $r_1=$60 $\mu$m and $r_2=$ 60, 80, 100 and 140 $\mu$m,
respectively. We measure the capacitance $C$ and conductance $G$
between the two gates using a cryogenic bridge and analyze its output
with a custom-made radio-frequency lock-in amplifier
\cite{Zhao.RSI.2022, Zhao.CPL.2022, Note1}.

Fig. 1(a) shows the $C$ and $G$ measured from the $r_1=r_2=$ 60 $\mu$m
sample. Both $C$ and $G$ decrease as we increase the magnetic field
$B$, owing to the magnetic localization where the 2DES conductance
$\sigma \propto (ne^2\tau)/m^{\star} (1+\omega_c^2\tau^2)$,
$m^{\star}$, $\omega_c$ and $\tau$ are the effective mass, cyclotron
frequency and transport scattering time of the electrons, respectively
\cite{Zhao.CPL.2022}. The $C$ and $G$ are finite at $\nu=1/2$ and 1/4
where the 2DES forms compressible composite Fermion Fermi sea. When
$\nu$ is an integer or a certain fraction such as 1/3 and 1/5, the
2DES forms incompressible quantum Hall liquids so that both $C$ and
$G$ vanish \footnote{The zero of $C$ and $G$ can be defined either by
  extrapolating their field dependence to $B=\infty$, or by their
  values at strong quantum hall states such as $\nu=1$. These two
  approaches are consistent with each other and the dash lines in
  Fig. 1(a) represent the deduced zero.}.

In all the above cases, the current is carried by \emph{transporting
  electrons}, so that $C$ has a positive dependence on $G$,
i.e. $C\propto G^{3/2}$, as shown in Fig. 1(b)
\cite{Zhao.CPL.2022}. Such a correlation discontinues when the WC
forms at very low filling factors $\nu\lesssim 1/5$, see the blue
shaded regions of Fig. 1(a). The vanishing conductance $G$ suggests
that the electrons are immovable, however, the surprisingly large
capacitance $C$ evidences that the WC hosts a current even surpassing
the conducting Fermi sea at $\nu=1/2$ and 1/4 at much lower magnetic
field! The phase transition between the WC and the liquid states are
clearly evidenced by spikes in $G$ (marked by solid circles in
Fig. 1(a)) and sharp raises in $C$. A developing minimum is seen in
$G$ at $1/5 < \nu <2/9$ (marked by the up-arrow) when $C$ has a
peak. This $G$ minimum develops towards zero and the $C$ peak
saturates when the solid phase is stronger (see black traces in
Fig. 3(a)). This is consistent with the reentrant insulating phase
\cite{Jiang.PRL.1990, Goldman.PRL.1990, Williams.PRL.1991,
  Li.PRL.1991, Chen.PRL.2004, Shayegan.Flatland.2006,
  Shayegan.PQHE.1998}.
 
It is important to mention that the 2DES in our devices is effectively
``isolated'' and we are merely transferring charges between different
regions within one quantum phase. Similar to the dielectric materials which also have no transporting electrons, the collective motion of all electrons,
i.e. the $k\to 0$ phonon mode of WC, can generate
\emph{polarization charges} and corresponding polarization current in response to the in-plane component of applied electric field. An infinitesimally small but ubiquitous
disorder pins the WC so that electrons can only be driven
out of their equilibrium lattice site by a small displacement
$\mathbf{x}$, as shown in Fig. 1(c). During the experiments, we use excitation
$V_{\text{in}} \simeq$ 0.1 mV$_{\text{rms}}$ and the measured WC capacitance is $\sim$ 0.15 pF at 13.5 T. The polarization charge
accumulated under the inner gate is $Q=C V_{\text{in}} \sim$ 100
$e$. The corresponding electron displacement at the boundary of the
inner gate, $|\mathbf{x}(r_1)| \simeq Q/(2\pi r_1ne)\sim 0.6$ nm, is
much smaller than the magnetic length $l_B=\sqrt{\hbar/eB}\sim 8$
nm, substantiating our assumption that the electrons vibrate
diminutively around their equilibrium lattice sites.
 
An ideal, disorder-free WC is effectively a perfect dielectric with
infinite permittivity, so that the device capacitance should be close
to its zero-field value $C_0 \sim$ 1 pF when 2DES is an excellent
conductor. We note that $C_0$ is consistent with the device geometry,
$\epsilon_0 \epsilon_{\text{GaAs}}\pi r_1^2/h \simeq$ 1.3 pF, where
$\epsilon_{\text{GaAs}}=12.8$ is the relative dielectric constant of
GaAs and $h \simeq $ 960 nm is the depth of 2DES. However, the
measured $C\sim 0.15$ pF in the WC regime is much smaller than
$C_0$. This discrepancy is likely caused by the friction-like disorder
which poses a pinning force $\simeq -\beta \mathbf{x}$ on the
electrons. When the crystal's inversion symmetry is broken,
i.e. $\mathbf{x}$ is non-uniform and $\mathcal{J}(\mathbf{x})$ is
finite, the electron-electron interaction generates a restoring force
$\simeq - a_0\mu_{ij} \mathcal{J}(\mathbf{x})$, where $\mu_{ij}$,
$a_0$ and $\mathcal{J}(\mathbf{x})$ are the elastic tensor, WC lattice
constant and the Jacobi matrix of $\mathbf{x}$, respectively. At the
low frequency limit, the WC is always at equilibrium and all forces
are balanced,
$e\mathbf{E}- a_0\mu_{ij} \mathcal{J}(\mathbf{x})-\beta\mathbf{x}=0$,
$\mathbf{E}$ is the total parallel electric field on the WC.

$\mathbf{E}$ is approximately zero under the metal gates, since the
gate-to-2DES distance $h$ is small. Therefore,
$\mathbf{x}$ decreases exponentially when the distance from the gate
boundary $d$ increases, $\mathbf{x}\propto \exp(-d/\zeta)$, where $\zeta=\mu a_0/\beta$ is the decay length. Deeply
inside the gates, electrons feel
neither parallel electric field nor net pressure from nearby
electrons, so that their displacement $\mathbf{x}$ remains
approximately zero. This region does not contribute to the capacitive
response, and the effective gate area reduces to about
$2\pi r_1 \zeta$ and $2\pi r_2 \zeta$ at the inner and outer gate, respectively. Because $r_1=r_2=$ 60 $\mu$m in Fig. 1(a), the experimentally measured $C \approx \epsilon_0 \epsilon_{\text{GaAs}}/h \cdot 2\pi r_1\zeta/2\simeq$ 0.15 pF at 13.5 T corresponds to a decay length $\zeta\simeq $ 6.7
$\mu$m. Interestingly, our result shows a linear dependence
$C\propto 1/B$ in Fig. 1(d), suggesting that $\beta\propto l_B^{-2}$
if we assume $\mu_{ij}$ is independent on $B$. Especially, the pinning
becomes infinitely strong,
i.e. $\beta\to \infty$, at the extreme quantum limit $l_B\to 0$.

The permittivity of a disorder-pinned WC is no longer infinitely
large, since a non-zero electric field $\mathbf{E}$ is necessary to
sustain a finite $\mathbf{x}$. If we assume $\mathbf{x}$ is a constant
in the ring area between the two gates, so that
$e\mathbf{E} = \beta\mathbf{x}$. The residual $\mathbf{E}$ can be
modeled as a serial capacitance
$C_{\text{WC}} \approx 2\pi ne^2/\beta \cdot [\ln(r_2/r_1)]^{-1}$ in
our device. We then measure different devices with $r_1$= 60 $\mu$m
and $r_2=60$, 80, 100 and 140 $\mu$m, and calculate the corresponding
$C_{\text{WC}}$ through
$C_{\text{WC}}^{-1}=C^{-1}-(r_1+r_2)/r_2\cdot C_{r_1=r_2}^{-1}$, see
Fig. 1(e). By fitting the linear dependence
$C_{\text{WC}}^{-1} \propto \ln(r_2/r_1)$, we estimate the pinning
strength $\beta$ to be about 1.3 $\times 10^{-9}$ and 1.1
$\times 10^{-9}$ N/m at $B=13.5$ and 12 T, respectively
\footnote{Alternatively, $C_{\text{WC}}$ can be modeled as a cylinder
  capacitor whose height equals the effective thickness of the 2DES,
  $Z_0 \approx 45 $ nm. The WC dielectric constant is
  $\epsilon_{\text{WC}}= (2\pi\epsilon_0 Z_0
  \partial(C_{\text{WC}}^{-1})/\partial \ln (r_2/r_1))^{-1} \approx 2
  \times 10^4$ at 13.5 T, consistent with previous reported value in
  ref. \cite{Li.SSC.1995}.}. Finally, assuming
$\mu_{ij}\approx \mu\cdot \delta_{ij}$, we can estimate the WC elastic
modulus $\mu\approx \beta\cdot \zeta /a_0 $. For example, $\mu$ is
about $1.6\times 10^{-7}$ N/m at 13.5 T.

\begin{figure}[!htbp] 
\includegraphics[width=0.45\textwidth]{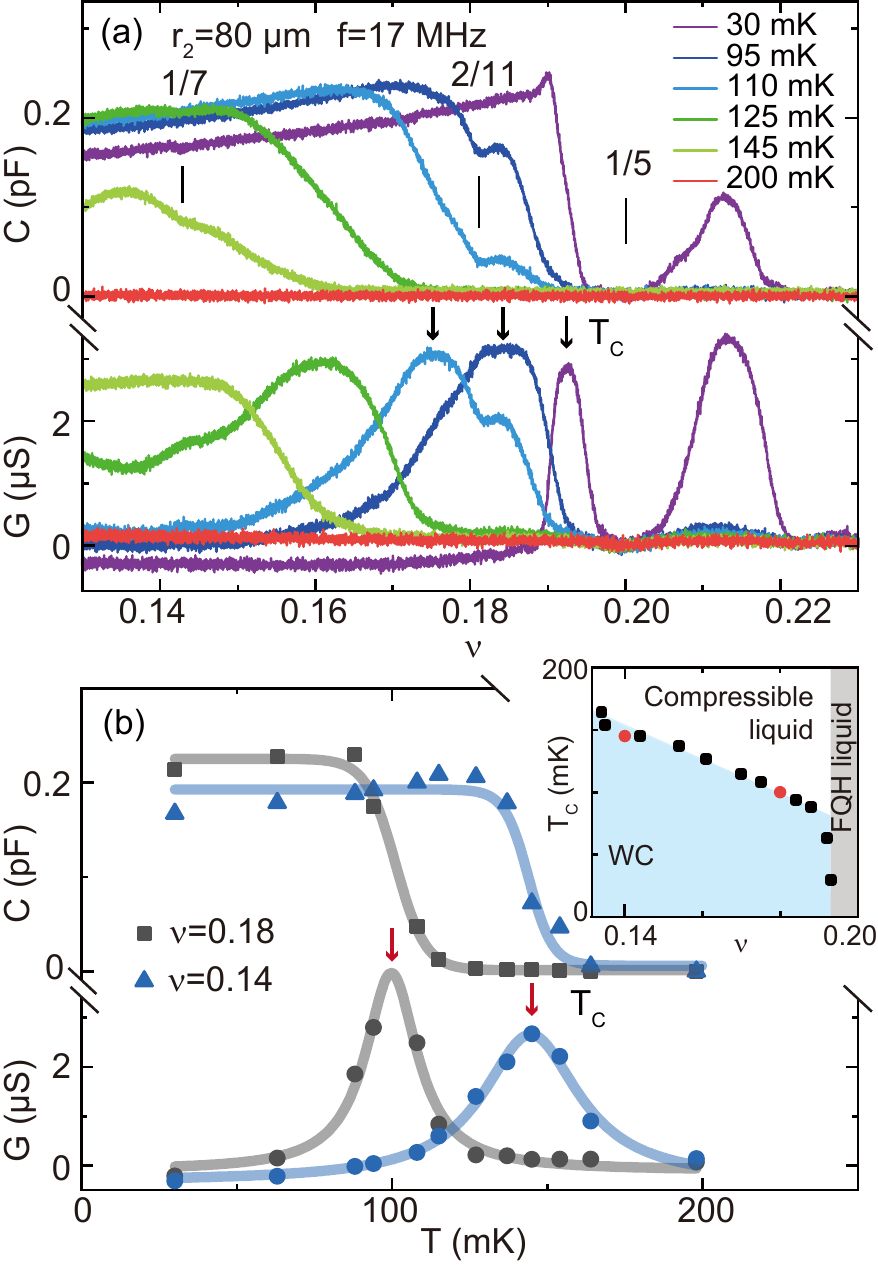} 
\caption{(color online) (a) $C$ and $G$ vs. $\nu$
  measured at various temperatures from the $r_2=$ 80 $\mu$m sample
  with 17 MHz excitation. (b) Summarized $C$ and $G$ vs. $T$ at
  $\nu=0.14$ and 0.18 from the panel (a) data. A critical temperature
  $T_c$ at certain $\nu$ is defined either as the temperature when $G$
  has a peak at $\nu$ in panel (a) or as the temperature when $G$
  vs. $T$ trace reaches maximum in panel (b); marked by the black and
  red arrows. The panel (b) inset summarizes the $T_c$ using the two
  equivalent definitions using black and red circles,
  respectively. The diagram can be separated into three different
  regions corresponding to the WC, the fractional quantum Hall (FQH)
  liquid and the compressible liquid.}
\end{figure}

\begin{figure*}[!htbp] 
\includegraphics[width=0.9\textwidth]{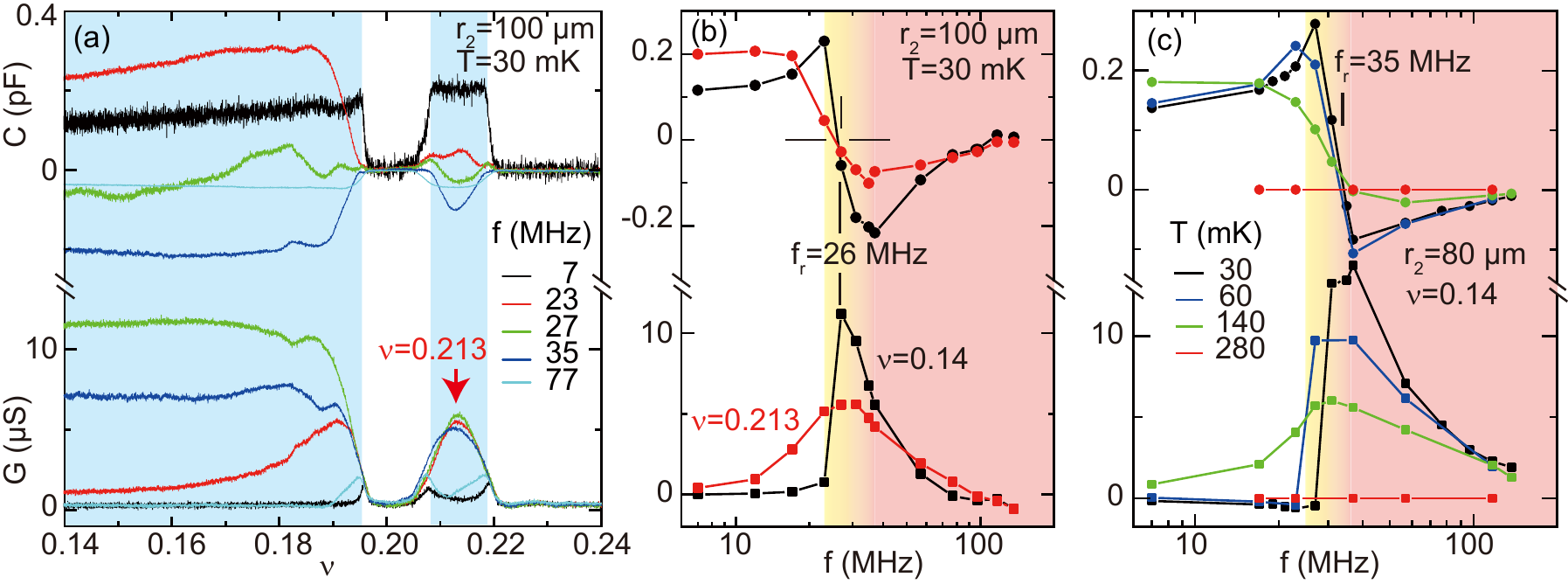} 
\caption{(color online) (a) $C$ and
  $G$ vs. $\nu$ taken from the $r_2$=100 $\mu$m sample using
  different excitation frequencies $f$. We see a violent change of $C$
  and $G$ at different $f$ in the blue region where the WC
  appears.  (b) The $C$ and $G$ vs. $f$ extracted from the panel (a)
  trace at $\nu=0.14$ and 0.213.  The resonance frequency $f_r$,
  defined as the frequency when $C$ changes its sign, is about 26
  MHz. (c) The $C$ and $G$ vs. $f$ at $\nu$ = 0.14 and different
  temperatures, data taken from the $r_2$=80 $\mu$m sample. The
  resonance disappears at $T\simeq$ 280 mK when $C$ and $G$ remain
  nearly zero.}
\end{figure*}
      
Fig. 2 reveals an intriguing temperature-induced solid-liquid phase
transition when the WC melts. Fig. 2(a) shows $C$ and $G$ taken from
the $r_2=80$ $\mu$m sample at various temperatures. At a certain
temperature, e.g. at $T\approx 110$ mK, $C\sim 0.2$ pF when the 2DES
forms WC at $\nu \lesssim 0.16$ and vanishes when it is a liquid phase
at $\nu \gtrsim 0.18$. $G$ has a peak at $\nu\simeq 0.175$ when $C$
vs. $\nu$ has the maximal negative slope, and it is small when the
2DES is either a WC at $\nu< 0.17$ or a liquid at $\nu> 0.19$
\footnote{We observe developing minimum at $\nu=1/7, 2/11$ during the
  solid-liquid phase transition, signaling that the fractional quantum
  Hall state emerges \cite{Pan.PRL.2002, Chung.PRL.2022}.}. At very
high temperature $T\gtrsim$ 200 mK, both $C$ and $G$ are close to
zero. In Fig. 2(b), we summarized $C$ and $G$ as a function of $T$ at
two different filling factors to better illustrate this solid-liquid
transition. At $\nu\simeq 0.14$, for example, $C$ is large and $G$ is
small at $T\lesssim 100$ mK when the WC is stable \footnote{$C$
  vs. $T$ has a slightly positive slope in the WC region, possibly due
  to the softening of disorder pinning.}, while both of them become
small at $T\gtrsim 200$ mK when the 2DES is a liquid. The $G$ has a
peak at a critical temperature $T_C$, marked by the red arrows, around
which the precipitous decrease of $C$ happens. Alternatively, $T_C$ at
a certain filling factor $\nu$ can be defined as the temperature when
the $G$ has a peak (black arrow in Fig. 2(a)) at $\nu$. We summarize
$T_C$ obtained using these two equivalent procedures in the Fig. 2(b)
inset with corresponding red and black symbols. $T_C$ has a linear
dependence on $\nu$ whose two intercepts are $T_C\simeq 340$ mK at the
extreme quantum limit $\nu=0$, and $\nu\simeq$ 1/4 at $T_C=0$ mK.

The Fig. 2(b) evolution can be qualitatively understood by the
coexistence of transport and polarization currents at the solid-liquid
transition. The large $C$ reduces to almost zero when the transport
current dominates over the polarization current. $G$ is a measure of
the 2DES's capacity to absorb and dissipate power. It is negligible if
either of these two currents dominates, since the polarization current
is dissipation-less and the dissipating transport current is difficult
to excite. $G$ becomes large when these two currents coexist nip and
tuck at intermediate $T$ when the excited polarization charge can be
just dissipated by the transport current.

The WC exhibits a resonance when we increase the excitation
frequency. In Fig. 3(a), the $C$ and $G$ measured from the $r_2= 100$
$\mu$m sample using different excitation frequencies change enormously
when the WC presents (blue shaded region). $G$ is almost zero and $C$
is large at $f\simeq 7$ MHz, and $G$ becomes finite and $C$ becomes
even larger at $f\simeq 23$ MHz. At slightly higher frequency 27 MHz,
$G$ reaches its maximum and $C$ drops to about zero. Further
increasing $f$, $G$ gradually declines while $C$ first becomes
negative at 35 MHz and then gradually approaches zero. The summarized
$C$ and $G$ vs.  $f$ at two certain fillings in Fig. 3(b), resembles
qualitatively a resonant behavior with resonance frequency
$f_r\simeq 26$ MHz (when $C=0$). Fig. 3(c) studies this resonance at
different temperatures. The data is taken from the $r_2\simeq$ 80
$\mu$m sample whose resonance frequency is about 35 MHz
\footnote{$f_r$ has no obvious dependence with sample geometry, which
  is about 35, 35, 26 and 29 MHz for samples with $r_2$ = 60, 80, 100,
  140 $\mu$m, respectively.}. The abrupt change of $C$ near $f_r$
becomes gradual and the $G$ peak flattens at higher temperatures. Both
$C$ and $G$ become flat zero at $T\gtrsim 280$ mK. It is noteworthy
that, as long as a resonance is seen, $f_r$ is nearly independent on
the filling factor (Fig. 3(b)) and temperatures (Fig. 3(c)). This is
consistent with another experimental study using surface acoustic wave
\cite{Drichko.PRB.2016}.

The resonance of WC is usually explained by the pinning
mode \cite{Fogler.PRB.2000, Ye.PRL.2002}. The resonance frequency is
related to the mean free path $L_T$ of the transverse phonon through
$L_T=(2\pi\mu_{t,cl}/neBf_r)^{1/2}$, where
$\mu_{t,cl}=0.245e^2n^{3/2}/4\pi \epsilon_0\epsilon_{\text{GaAs}}$ is
the classical shear modulus of WC.  $f_r=26$ MHz corresponds to
$L_T \simeq$ 3.2 $\mu$m, very similar to $\zeta \simeq 6.7$ $\mu$m in
our Fig. 1(c) discussion. This is justifiable because both $L_T$ and
$\zeta$ describe the length-scale within which the collective motion
of WC is damped/scattered by the random pinning potential.

Before ending the discussion, we would like to highlight the puzzling
"half-dome" structure of the resonance. $G$ has
a regular-shaped resonance peak, i.e. $G$ decreases gradually on both
sides of $f_r$, when either the WC is weak ( $\nu\simeq 0.213$ in
Fig. 3(b)) or the temperature is high ($T\simeq 140$ mK in
Fig. 3(c)). Surprisingly, the resonance peak becomes quite peculiar when the
WC is strong at $\nu\simeq 0.14$ and $T\simeq 30$ mK. $G$
gradually decreases from its peak at $f_r$ on the high frequency side
$f>f_r$, while it vanishes instantly when the frequency is lower than
$f_r$, resulting in a "half-dome" $G$ vs. $f$ trace. Meanwhile, the
$C$ increases by $\sim 2$ times and then abruptly changes
to negative at $f_r$. This anomalous "half-dome" feature is seen in
all of our devices as long as the WC is strong and
temperature is sufficiently low, suggesting a threshold frequency for
the power dissipation.

In conclusion, using the extraordinarily high-precision capacitance
measurement technique, we investigate the dynamic response of WC
systematically. From the quantitative results and using a simple
model, we can study several physical properties of the WC such as
elastic modulus, dielectric constant, pinning strength, etc., and
discover a puzzling "half-dome" feature in the resonance peak. Our
results certainly shine light on the study of WC and provides new
insight on its dynamics.

\begin{acknowledgments}
  
  We acknowledge support by the National Nature Science Foundation of China (Grant No. 92065104 and 12074010) and the National Basic Research Program of China (Grant No. 2019YFA0308403) for sample fabrication and measurement.  This research is funded in part by the Gordon and Betty Moore Foundation’s EPiQS Initiative, Grant GBMF9615 to L. N. Pfeiffer, and by the National Science Foundation MRSEC grant DMR 2011750 to Princeton University.  We thank L. W. Engel, Bo Yang and Xin Lin for valuable discussion.

\end{acknowledgments}

%\nocite{*}
\bibliography{bib_full}

\newpage

\renewcommand*{\thefigure}{S1}

\section{Supplementary Materials}

\subsection{Samples}

The sample we studied is made from a GaAs/AlGaAs heterostructure wafer
grown by molecular beam epitaxy. A 70 nm-wide GaAs quantum well is
bound by AlGaAs spacer-layers and $\delta$-doped layers on each side,
and locates $h \simeq$ 960 nm below the sample surface. The as-grown
density of the 2DES is $n\simeq4.4\times10^{10}$ cm$^{-2}$, and its
mobility at 300 mK is $\mu\simeq $ 17 $\times10^6$
cm$^2$/(V$\cdot$s). Our sample is a 2 mm $\times$ 2 mm square piece
with four In/Sn contacts at each corner. The contacts are grounded through a resistor to avoid signal leaking. We evaporate concentric,
Au/Ti front gate pair G1 and G2 using standard lift-off process, whose
outer and inner radius is $r_1$ and $r_2$, respectively. We deposit a
20 nm thick Al$_2$O$_3$ layer between the two gates to prevent them
from shorting with each other. The four outer-gates are merged into one piece so that
the area of the outer gate G2 is much larger than the inner gate G1.

\subsection{Capacitance Measurement Setup}

\begin{figure*}[!htbp]
	\includegraphics[width=0.9\textwidth]{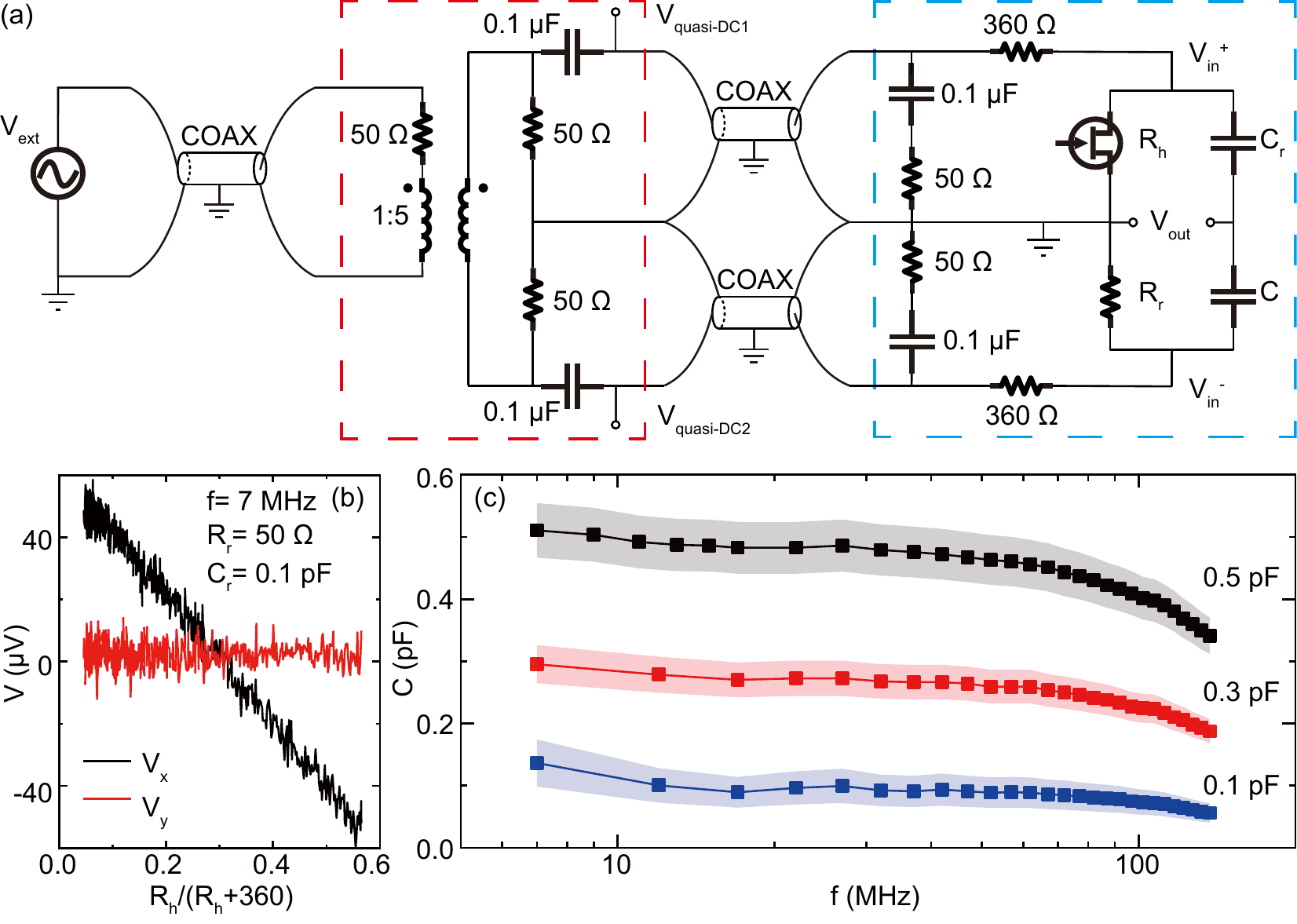}
	\caption{(color online) 
		(a) Circuit diagram of measurement bridge with 50
		$\Omega$ impedance match networks. 
		(b) The $V_{\text{X}}$ and $V_{\text{Y}}$ from a typical ``V-curve'' procedure. $C$ is about 0.25 pF from the balance condition Eq. (5).
		(c) The calibration results, by measuring
		commercial capacitors with different frequencies.
	}
\end{figure*}

The capacitance and conductance response is measured with a cryogenic
bridge similar to refs. \cite{Zhao.RSI.2022, Zhao.CPL.2022}.

The kernel of the bridge consists four devices, $R_{\text{h}}$,
$R_{\text{r}}$, $C_{\text{r}}$ and $C$, as shown in Fig. S1(a). $C$ is
the capacitance of sample. We change the value of $R_{\text{h}}$ to
reach the balance condition 
\begin{equation}
	\frac{C}{C_{\text{r}}}=\frac{R_{\text{h}}}{R_{\text{r}}}.
\end{equation}
The bridge output $V_{\text{out}}$ is minimum at the balance
condition, from which we calculate the $C$. This is the so-call
``V-curve'' procedure, see refs. \cite{Zhao.RSI.2022, Zhao.CPL.2022}
for more information.

In order to expand the allowed bandwidth of the excitation frequency,
we add an impedance match network to the input of the bridge, shown as
the Fig. S1(a). $V_{\text{ext}}$ is the signal source with 50 $\Omega$
output impedance. $V_{\text{ext}}$ drives a signal splitter box (the
red dashed box) located at the top of the dilution refrigerator
through a $\sim$2 m-long semi-rigid coaxial cable. The box input is a
1:5 transformer in series with a 50 $\Omega$ resistor. The transformer
output drives two serial connected 50 $\Omega$ resistors
differentially. The differential signals are transmitted to the
cryogenic sample holder (the blue dotted box) by two rigid coaxial
cables of $\sim$2 m length. Another pair of impedance matching 50
$\Omega$ resistors are added at the input of the cryogenic bridge, and
the 360 $\Omega$ resistors are chosen by balancing the competition
between the performance and heating. The characteristic impedance of all coaxial cables in the work is 50 $\Omega$. 

The low-frequency signals $V_{\text{quasi-DC1}}$ and
$V_{\text{quasi-DC2}}$ used to measure the value of $R_{\text{h}}$ and
$R_{\text{r}}$, respectively. The 0.1 $\mu$F capacitors are used to
separate the high-frequency excitation signals and the quasi-DC
signal.

The output $V_{\text{out}}$ is approximately 
\begin{equation}
	V_{\text{out}} \propto S \cdot (\frac{R_{\text{h}}}{360+R_{\text{h}}}-\frac{C}{C_{\text{r}}} \cdot \frac{R_{\text{r}}}{360+R_{\text{r}}}) \cdot V_{\text{ext}}.
\end{equation}
$S$ can be obtain from the ``V-curve'' procedure by linear fitting the $V_{\text{X}}$ vs. $R_{\text{h}}/(360+R_{\text{h}})$, as shown in Fig. S1(b). $V_{\text{X}}$ and $V_{\text{Y}}$ are the orthogonal component of $V_{\text{out}}$, 

\begin{numcases}{}
	V_{\text{X}}=|V_{\text{out}}|\cdot \cos(\theta),\\
	V_{\text{Y}}=|V_{\text{out}}|\cdot \sin(\theta),
\end{numcases}
where $\theta$ is the phase of $V_{\text{out}}$. We can derive the
value of $C$ using Eq. (2) and (3). The new balance condition of the
revised bridge is 
\begin{equation}
	\frac{C}{C_{\text{r}}}=\frac{R_{\text{h}}}{R_{\text{r}}} \cdot\frac{360+R_{\text{r}}}{360+R_{\text{h}}},
\end{equation}
where the $V_{\text{X}}=0$.

Note that the capacitance $C$ and the conductance $G$ of sample lead to the orthogonal component $V_{\text{X}}$ and $V_{\text{Y}}$, respectively. Therefore, the $G$ can be obtained from Eq. (2) and (4) by replacing $C/{C_{\text{r}}}$ with $G/2\pi f {C_{\text{r}}}$, where $f$ is the excitation frequency.

Fig. S1(c) shows our calibration measurement using
different excitation frequencies. The data is almost flat from 7 to
$\sim$100 MHz. The measured capacitance begins to
decline slowly above $\sim$100 MHz, possibly due to the parasitic inductance of bonding wires.

\end{document}